# Electronic structure and local distortions in epitaxial ScGaN films

S.M. Knoll,[a] M. Rovezzi,[b] S. Zhang,[a] T.B. Joyce[c] and M.A. Moram[d]

[a] Dept. Materials Science & Metallurgy, University of Cambridge, 27 Charles Babbage Rd., Cambridge, CB3 0FS, UK.

[b] European Synchrotron Radiation Facility, BP220, 38043 Grenoble, France

[c] Department of Materials Science and Engineering, University of Liverpool, Ashton St, Liverpool, L69 3GH, UK

[d] Dept. Materials, Imperial College London, Exhibition Road, London, SW7 2AZ, UK

## Abstract

High energy-resolution fluorescence-detected X-ray absorption spectroscopy and density functional theory calculations were used to investigate the local bonding and electronic structure of Sc in epitaxial wurtzite-structure $Sc_xGa_{1-x}N$ films with $x \leq 0.059$. Sc atoms are found to substitute for Ga atoms, accompanied by a local distortion involving an increase in the internal lattice parameter $u$ around the Sc atoms. The local bonding and electronic structure at Sc are not affected strongly by the strain state or the defect microstructure of the films. These data are consistent with theoretical predictions regarding the electronic structure of dilute $Sc_xGa_{1-x}N$ alloys.

## 1. Introduction

The III-nitride semiconductors InN, GaN and AlN are used widely in optoelectronic and high power electronic devices. However, internal quantum efficiencies remain relatively low in the ultraviolet [1][2] and green spectral regions. This has motivated the search for alternative alloying elements to expand the functionality of the existing III-nitrides. Alloys with Sc are of particular interest in this regard.

Unlike other transition metal nitrides, ScN is a semiconductor with a direct band gap of 2.1 eV [3][4] and an indirect gap of 0.9 eV [5]. Although ScN has a rock-salt structure, it is expected to form stable wurtzite-structure $Sc_xGa_{1-x}N$ alloys up to $x = 0.27$, whilst retaining a direct band gap [6]. In this respect, Sc is different from other transition metals, which have low solubilities in GaN [7]. Density functional theory (DFT) calculations have found that in the dilute Sc regime ($x < 0.27$), the wurtzite $Sc_xGa_{1-x}N$ structure experiences a gradual reduction in the $c/a$ ratio with increasing $x$, accompanied by a gradual increase in the internal $u$ parameter [6]. In particular, this increase in $u$ is mainly manifested as a local distortion around the Sc atoms, which can be visualised as a transition moving from the four-fold tetrahedral ($T_d$) coordination of the wurtzite phase towards a five-fold, trigonal-bipyramidal ($B_k$) coordination as found in hexagonal BN (h-BN) [8]. Previous experimental studies with transmission electron microscopy (TEM) and high-resolution X-ray diffraction (HRXRD) have found indications of the presence of such a local distortion to occur, and shown that there is no detectable phase segregation for low Sc-content $Sc_xGa_{1-x}N$ alloys below at least $x = 0.17$ [9][10]. However, XRD is only sensitive to long-range ordering, and may not detect nano-crystalline precipitation, whilst even HRTEM is unable to distinguish between the structural differences around Sc or Ga due to their relatively similar atomic masses and low concentrations.

X-ray absorption spectroscopy (XAS) is an element and spin/orbital angular momentum selective technique with sensitivity to both crystalline and amorphous phases. It permits to probe the local electronic structure and geometric arrangement of a given element in a material [11]. In this study, we have used XAS to provide information about the local coordination environment, bonding and electronic properties of Sc in $Sc_xGa_{1-x}N$. Our data are compared to the corresponding theoretical predictions for this technologically promising material.

## 2. Experimental

### 2.1 Experimental techniques

Epitaxial [0001]-oriented $Sc_xGa_{1-x}N$ films with Sc contents up to $x = 0.059$ were grown on [0001]-oriented GaN-on-sapphire substrates using molecular beam epitaxy (MBE) with $NH_3$ as a nitrogen source. The GaN-on-sapphire pseudo-substrate was grown using metal-organic vapour phase epitaxy (MOVPE) [12]. Details of the $Sc_xGa_{1-x}N$ growth conditions have been reported previously [9][13]. HRXRD was performed using a Panalytical X-Pert Pro MRD machine in the triple-axis configuration using a four-bounce asymmetric Ge(220) monochromator and a triple-bounce analyser. Cross-sectional transmission electron microscopy (TEM) analysis was also performed on all samples. The HRXRD rocking curves of the GaN 0002 reflection showed full-width at half-maxima (FWHM) of 240 ± 10 arcsec and the TEM data indicated that all $Sc_xGa_{1-x}N$ films were epitaxial. A detailed analysis of the film microstructure will be reported elsewhere [14]. The scandium content of the films was determined accurately by comparing the relative fluorescence intensities of the Sc K$\alpha_1$ line using a stoichiometric ScN reference film grown by MBE under similar conditions, with a correction for the film thicknesses and densities.

The XAS experiments at the Sc K-edge were performed on the ID26 beamline [15] of the European Synchrotron Radiation Facility. The X-ray beam, linearly polarised in the horizontal plane, was produced by three coupled undulators and monochromatised using a pair of cryogenically-cooled Si(111) crystals. Rejection of the harmonics was achieved via a Si-coated mirror run at 3.5 mrad glancing angle and the focusing was obtained with two adaptive Si-coated mirrors (bimorph piezoelectric systems) run in the Kirkpatrick-Baez geometry. At the Sc K-edge (4488.6 eV), this configuration permits an incident beam of horizontal width ~600 μm and vertical height ~50 μm to be obtained on the sample, with a flux of ~$10^{13}$ photons/s and an energy resolution of 0.64 eV FWHM. The fluorescence emitted by the sample was monochromatised with a Johann-type spectrometer equipped with five analysers and run in the vertical Rowland geometry at a scattering angle of 90° [16]. This permits the collection of high energy-resolution fluorescence-detected (HERFD) XAS spectra. HERFD-XAS spectra were collected at the K$\alpha_1$ line using five Si analysers at the (311) reflection (Bragg angle ~67°), obtaining an outgoing energy resolution of 0.89 eV FWHM. The HERFD-XAS data was collected in grazing incidence conditions (~5° incidence) with the polarisation vector parallel and perpendicular to the $c$-axis, that is, with the sample vertical and horizontal to the scattering plane, respectively. The extended X-ray absorption fine-structure (EXAFS) signal $\chi(k)$ was extracted using the Viper program [17], with the threshold energy $E_0$ set at the maximum derivative and a smoothing spline used to fit the post-edge background. X-ray absorption near edge structure (XANES) spectra were normalised to unity by averaging the post-edge function.

### 2.2 Theoretical methods

The calculations of the unoccupied electronic projected density of states (PDOS) of ScN and the dilute ScGaN systems performed in this study are based on density functional theory (DFT) within the generalised gradient approximation [18] using linearised augmented plane waves implemented in the WIEN2k code [19]. Supercells were constructed for both wurtzite $Sc_1Ga_{15}N_{16}$ and $Sc_2Ga_{14}N_{16}$ (2×2×2 duplication of the wurtzite GaN unit cell, 32 atoms) The $Sc_1Ga_{15}N_{16}$ supercell corresponds to a composition of 6.25% Sc and are therefore suitable for direct comparison with the $x = 0.059$ $Sc_xGa_{1-x}N$ sample. It was not practical to simulate the PDOS of lower Sc content $Sc_xGa_{1-x}N$ alloys due to the high computational demands associated with the larger supercells required. The lattice parameters and internal coordinates of each supercell were firstly relaxed to find the global energy minima [6]. Then, a $1s$ core hole was introduced at the probed Sc atom, whose PDOS was then calculated by WIEN2k. XANES spectra were simulated using the FDMNES code [20][21] employing the relaxed $Sc_1Ga_{15}N_{16}$ supercell. The Green formalism (multiple scattering) on a muffin-tin potential has been applied, with a cluster radius of 10 Å. The convolution was performed with an energy-dependent Lorentzian function to correctly account for the initial and final states energy broadening and with a Gaussian function of constant FWHM (0.6 eV), to account for instrumental broadening.

## 3. Results & discussion

### 3.1 EXAFS studies

The effect of alloy composition on the local coordination environment of Sc was investigated by looking at a series of $Sc_xGa_{1-x}N$ samples with increasing Sc content. A complementary HRTEM study [14] revealed that Sc appears as a solid solution without any evidence of compositional clustering; however a marked increase in the defect density, especially in the form of basal plane stacking faults, was observed at increasing compositions. Fig. 1 (a) shows a plot of the $k^2$-weighted EXAFS data, $\chi(k)$, with the polarisation perpendicular and parallel to the wurtzite $c$-axis. This permits to resolve the in-plane and out-of-plane bonding geometry around Sc, respectively. In fact, the EXAFS signal depends on polarisation following the simple formula:

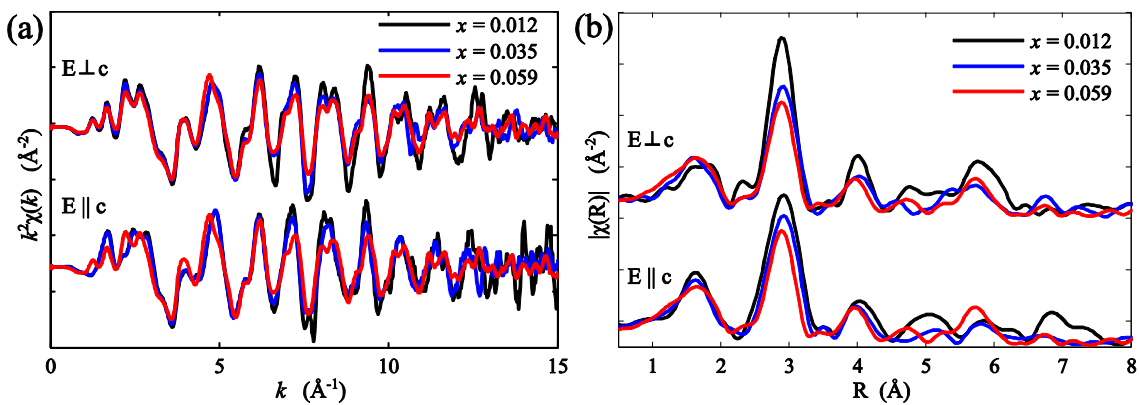

Fig. 1. Compositional evolution of EXAFS data for $Sc_xGa_{1-x}N$ samples with increasing $x$, for two perpendicular orientations. Panel (a) shows $k^2$-weighted spectra in $k$-space and (b) in Fourier transformed R-space.

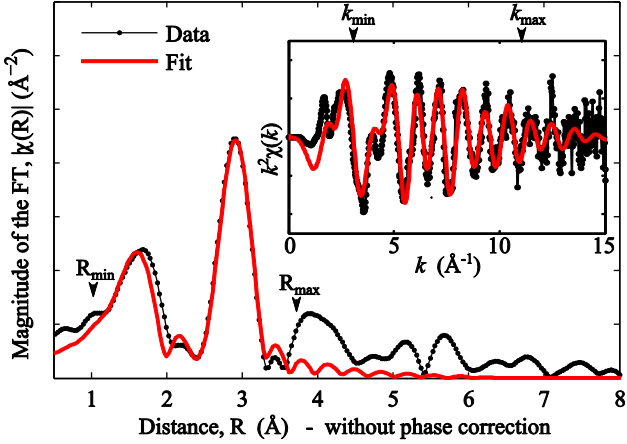

Fig. 2 EXAFS data and relative fit for $Sc_xGa_{1-x}N$ with $x = 0.059$, shown in R-space and in $k$-space (inset). The fitting was performed in R-space.

$$\chi(k)_{pol} = 3 \sum_i \cos^2(\theta_i)\chi(k)$$

where $\theta_i$ is the angle between the polarisation vector, $\epsilon$, and the distance vector of the scattering path considered, $R_i$. Qualitatively there is an amplitude reduction of the damping signal with the increasing Sc concentration, suggesting that the majority of Sc atoms remain in the same environment, while the structural disorder increases. Fig. 1 (b) shows the magnitude of the Fourier transformed (FT) EXAFS signal (R-space) versus Sc content. In a simple picture, the amplitude of the FT can be seen as a radial distribution function around the absorbing Sc, where the R scale is not correct due to the lost phase. The first two peaks arise mainly from single scattering events, the first one from Sc–N (4 nearest neighbours in wurtzite), while the second peak is mainly due to Sc–Ga (12 next nearest neighbours in wurtzite). The peaks at higher frequencies are due to mixed single and multiple scattering events and are discussed later. With the increasing Sc content, a reduction in the signal amplitude is observed in the second coordination shell. This can be ascribed to the increasing disorder in the sample, and is in agreement with the previous study of the microstructure by HRTEM [14]. Thus a quantitative fit was performed to investigate the amplitude reduction in detail.

Quantitative analysis was carried out using the IFEF-FIT-ARTEMIS programs [22][23], using the $k^2$-weighted EXAFS data in the range of $k = [3, 11]$ Å$^{-1}$, as indicated on the spectra in Fig. 2 by $k_{min}$ and $k_{max}$. The fits were performed in Fourier-transformed space, $\chi(R)$, over the range of $R = [1, 3.6]$ Å. Forward and reverse Fourier transforms over these ranges were performed using Hanning windows, with slope parameters of $dk = 1$ and $dR = 0.5$, respectively. The spectra were fitted using a two-shell model, corresponding to Sc–N ($R_N$) and Sc–Ga ($R_{Ga}$) coordination shells. The free parameters used in the fitting model were: the interatomic distances ($R_i$) plus the Debye-Waller factor for the cation shell, $\sigma_{Ga}^2$. The common amplitude reduction factor was kept constant at $S_0^2 = 0.935$, that is, the one calculated for GaN. This choice, with the one of constraining $\sigma_N$ to $\sigma_{Ga}$ was dictated by the need to remove the numerical correlation between the amplitude variables. The shift in threshold energy, $\Delta E_0$, was kept constant to the one fitted for the most dilute sample, that is $\Delta E_0^{0.012} = 0(1)$ eV.

Table 1 Summary of the fitting parameters used to model the EXAFS spectra. Uncertainties in the final Fig. are indicated in brackets.

| Sc content (x) | Sc–N bond length (Å) | | Sc–Ga bond length (Å) | | $\sigma_{Ga}$ (Å$^{-2}$) |
|---|---|---|---|---|---|
| | E ⊥ c | E ∥ c | E ⊥ c | E ∥ c | |
| 0.012 | 2.046(9) | 2.075(9) | 3.206(3) | 3.218(3) | 0.0058(4) |
| 0.035 | 2.043(7) | 2.073(3) | 3.214(3) | 3.225(3) | 0.0072(4) |
| 0.059 | 2.05(1) | 2.08(1) | 3.209(4) | 3.220(4) | 0.0085(5) |

The fitted EXAFS spectrum and respective FT for $x = 0.059$ is shown in Fig. 2. The quantitative results for the best fit are reported in Table 1. The quality of the fit is demonstrated by low R-factors (< 8%), propagated in the reported error bars (including correlations between variables). We observe a distortion in the first Sc–N coordination shell with smaller effect on Sc–Ga consistent with a localised increase in the internal lattice parameter $u$ (i.e. a deviation from perfect $T_d$ bonding environment around the Sc atoms). There is a substantial deformation in the Sc–N bond shell, with increases in the bond lengths of 5.3 ± 0.1% and 6.4 ± 0.5%, in and perpendicular to the basal plane, respectively. The effect is much reduced for the Sc–Ga shell, with bond elongations between 0.78 ± 0.01% (in-plane) and 1.42 ± 0.01% (out-of-plane). This result is consistent with other studies on transition metal doped GaN (e.g. Mn-doped GaN), where similar localised distortions are observed around the transition metal atoms [24]. The decreasing amplitude of the EXAFS signal with Sc content was well fitted by the increasing $\sigma_{Ga}$, consistent with the expected increase in the inherent random alloy disorder.

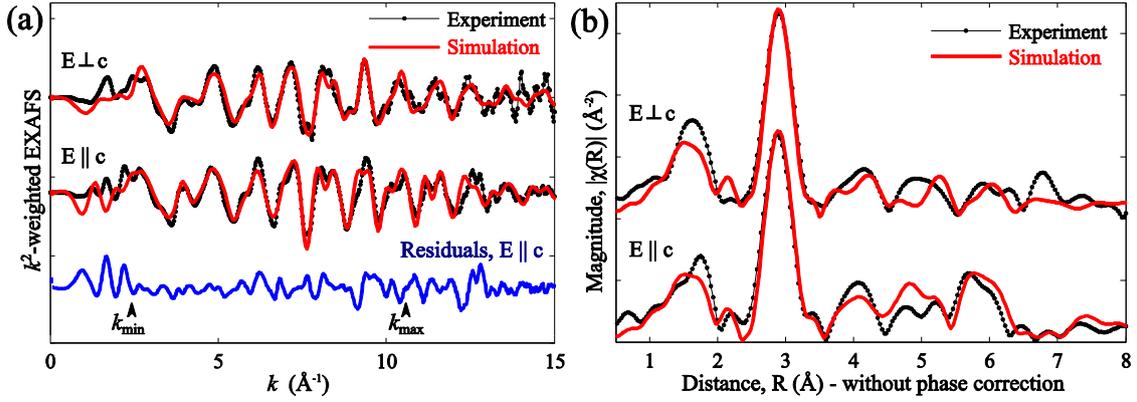

Fig. 4 (a) $k^2$-weighted EXAFS data and relative fits for orientations perpendicular and parallel to the wurtzite $c$-axis. The residual plot of the latter pair of curves is shown below. (b) FT experimental data and relative fits in R-space for both polarisations.

This strong out-of-plane deformation is also observed in our DFT calculations of the relaxed $Sc_1Ga_{15}N_{16}$ structure. The supercells have reduced $c/a$ lattice parameter ratios compared to those of relaxed $Ga_{15}N_{16}$ supercells. There is almost no change in the average internal parameter $u$, at the Ga atoms in the $Sc_1Ga_{15}N_{16}$ supercell compared to the average $u$ value at the Ga atoms in a relaxed $Ga_{16}N_{16}$ supercell ($u = 0.377$) [10]. However, there is a local increase in $u$ at the Sc atom ($u = 0.395$), accompanied by an increase in the metal-nitrogen bond lengths, as illustrated in Fig. 3. Further details regarding our calculated lattice parameters and $u$ values are given in Zhang et al [6].

To extend the analysis of the EXAFS data to higher coordination shells not considered in the previous model and also include the possibility to fit a minority contribution from additional crystallographic phases, we have used an EXAFS simulation approach, based on the relaxed $Sc_1Ga_{15}N_{16}$ supercell via DFT (Fig. 3). This approach is justified as the large number of scattering paths would considerably increase the number of free parameters, preventing to get accurate quantitative results. It has been demonstrated to be a valid approach in a similar system [25] [26]. In addition to the dilute $Sc_1Ga_{15}N_{16}$ model (near $T_d$ coordination of Sc), it is possible that a minority fraction of the Sc atoms are incorporated in the octahedral ($O_h$) environment found in rock-salt ScN, thus also this structure is taken into account in the EXAFS simulations. For each structure relaxed by DFT, the EXAFS spectra were simulated using the FEFF9.6 code [27] using full multiple scattering paths up to a length of R = 10 Å. The damping in the experimental EXAFS signal was accounted for using a correlated Debye model with the GaN Debye temperature of 600 K [28], which simulates the effects of thermal disorder. The effects of structural disorder were reproduced by the introduction of an additional common Debye-Waller factor ($\sigma_{Ga}$) ranging from 0.0015 – 0.0045 Å$^{-2}$. The use of smaller $\sigma_{Ga}$ factors than the ones found for the previous two-shell model is justified by the fact that for such large cluster (> 350 atoms from the absorbing Sc) the average structural disorder is smaller.

A linear combination of the simulated EXAFS spectra is then fitted to the experimental data over a $k$-range of $k$ = [2.5, 10.5] Å$^{-1}$, using the least-squares approach. This permits to obtain a quantitative measure of the fraction of Sc atoms in each coordination environment. Numerical results of the fit results are provided in Table 2.

Table 2 Results of linear combination of fits between simulated wurtzite $Sc_1Ga_{15}N_{16}$ and rock-salt ScN spectra, uncertainties in the last digit are given in brackets. A statistical increase in ScN fraction is found for $E \parallel c$, although the same trend could not be identified for $E \perp c$, because for a Sc content of 0.012 and 0.035 the use of ScN phase did not bring a statistical improvement of the fit quality.

| | $E \perp c$-axis | | $E \parallel c$-axis | |
|---|---|---|---|---|
| Sc content ($x$) | $Sc_1Ga_{15}N_{16}$ | ScN | $Sc_1Ga_{15}N_{16}$ | ScN |
| 0.012 | 1 | 0 | 0.96(3) | 0.04(3) |
| 0.035 | 1 | 0 | 0.92(2) | 0.08(2) |
| 0.059 | 0.94(3) | 0.06(3) | 0.87(3) | 0.13(3) |

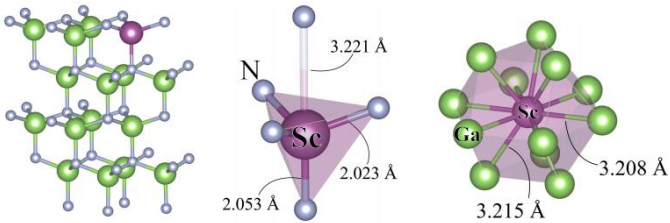

Fig. 3 Local bonding at the metal atoms in $Sc_1Ga_{15}N_{16}$, as predicted by density functional theory calculations, showing the four-fold coordination of Sc by N (1$^{st}$ coordination shell), and the 12-fold coordination of Sc by Ga (2$^{nd}$ coordination shell).

The combined simulated spectra were found to be in good agreement with the experiment for the most dilute case ($x = 0.012$), for data collected with the X-ray electric field vector polarised either parallel or perpendicular to the $c$-axis of the wurtzite crystal structure (Fig. 4). The decreased experimental amplitude with Sc content was again successfully reproduced by simply increasing $\sigma_{Ga}$.

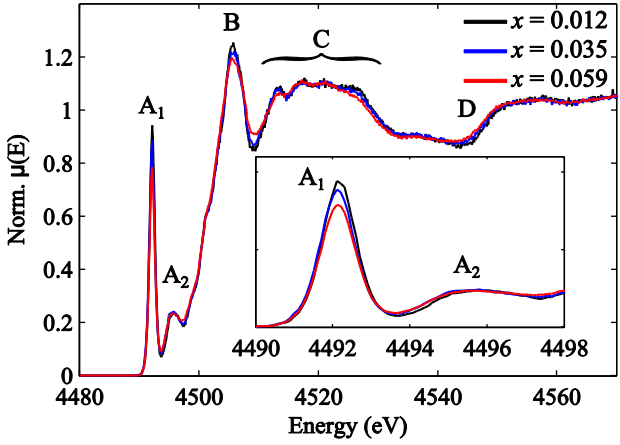

Fig. 5 HERFD-XAS spectra of $Sc_xGa_{1-x}N$ films at $x = 0.012$, $x = 0.035$ and $x = 0.059$, obtained from averaging spectra taken with the polarisation $E \perp c$ and $E \parallel c$ to eliminate dichroic effects (described later). The inset shows the overlaid $A_1$ and $A_2$ pre-edge peaks.

Overall, we observe that 95 ± 5 % of Sc is in a $T_d$ coordination environment for a Sc content of $x = 0.012$, possibly decreasing to 90 ± 10 % of Sc for higher contents up to $x = 0.059$. Resolving into the two orthogonal polarisations, we find that there is a statistically significant increase in the ScN fraction with increasing Sc content with $E \parallel c$, from 3.9 ± 2.6 % to 13.1 ± 2.6 %, whereas no such trend was found in the perpendicular orientation. These results show that while there are no measurable changes to the Sc coordination in the basal plane, there appears to be a change in the coordination along the $c$-axis. In this direction, the Sc atoms are moving from a near-$T_d$ configuration (with a single Sc–N bond along the $c$-axis) towards an $O_h$ configuration (with two Sc–N bonds along the $c$-axis). The resultant coordination can be envisaged as moving from a four-fold towards a five-fold coordinated structure ($B_k$). This is the same effect that would be seen for a decrease in the local $u$ parameter around the Sc atoms, in agreement with the previous EXAFS fits.

### 3.2 HERFD-XAS studies

The electronic structure of Sc in $Sc_xGa_{1-x}N$ was investigated by HERFD-XAS. Fig. 5 shows the experimental spectra derived as a weighted average of the two polarised XAS spectra $[2 \times (E \perp c) + (E \parallel c)] / 3$ in which dichroic effects are eliminated. All three samples investigated show the same main features, consisting of two well-defined peaks in the region before the main absorption edge (labelled $A_1$ and $A_2$), and several peaks in the fine structure region (labelled B–D, respectively). Both the number and position of the peaks remained constant with increasing composition, indicating that the average electronic and bonding environments of the Sc atoms in each sample are similar. However, we can observe two trends with increasing Sc content. Firstly, there is a slight reduction in intensity of the first pre-edge peak ($A_1$). Secondly, an overall loss in oscillation amplitude in the post-edge region is observed as the Sc content increases. These observations are likely to be attributed to an increased degree of structural disorder in the higher Sc content sample, as previously found by EXAFS.

The post-edge structure of the HERFD-XAS spectrum is predominantly affected by the local geometric structure around the absorbing atom, and to a lesser extent by the atomic structure of the absorbing element. This is illustrated by the resemblance of the fine structure observed in dilute ScGaN samples (Fig. 5) and other dilute systems such as MnGaN [24] and MnZnO [29], in which another transition metal substitutes for the cation of the wurtzite host lattice. In each case the transition metal is in tetrahedral coordination, giving rise to similar features in the post-edge regions B–D of the HERFD-XAS spectrum.

The experimental data was compared to simulated XANES spectra calculated by FDMNES [20][21]. Fig. 6 shows the experimental and simulated spectra for both polarisations, normalised over the same energy range. The main spectral features A–D are well reproduced by the FDMNES simulations. In particular, the presence of the two pre-edge peaks $A_1$ and $A_2$ is observed for both polarisations. The most striking feature of the $Sc_xGa_{1-x}N$ spectrum is the intense $A_1$ peak, with an intensity as high as 80% of the absorption edge jump. Such strong pre-edge peaks are expected to arise for transition metals bonded in tetrahedral environments due to electric dipole transitions to the $p$-component of $d$–$p$ mixed orbitals [30]. This is a strong indication that Sc is present in a local $T_d$ bonding environment, rather than $O_h$ as found in rock-salt ScN. To support this argument, we have performed PDOS calculations for Sc in different local bonding environments (Fig. 7). Our experimental data were collected at the Sc K-edge, which corresponds mainly to Sc $1s\rightarrow 4p$ dipole transitions. The imaginary part of the scattering amplitude is proportional to the empty $p$ projected density of states of the absorbing atom. Therefore it is possible to utilise DFT calculations of the PDOS to interpret features in the experimental XAS spectra [18]. The PDOS were calculated for Sc in an $O_h$ environment, Sc in a trigonal-bipyramidal environment (the $B_k$ structure, similar to $h$-BN [8]), and Sc in a $T_d$ ($Sc_1Ga_{15}N_{16}$).

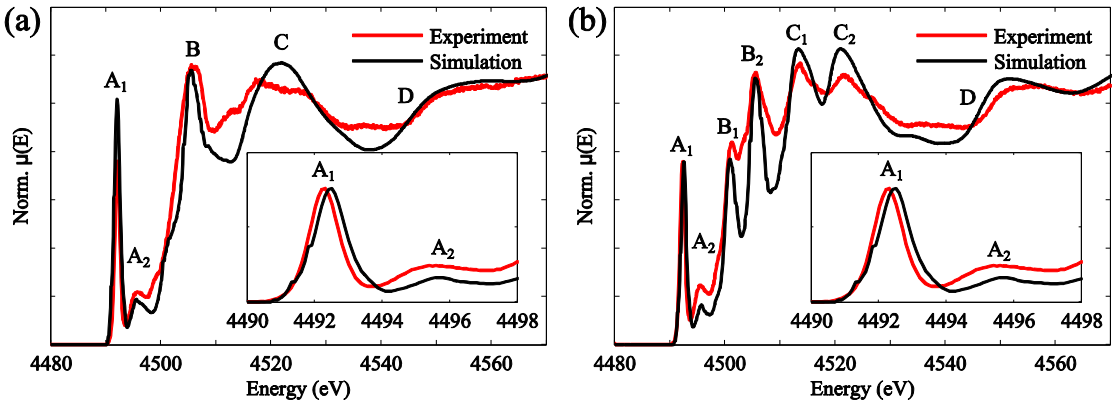

Fig. 6 Comparison of the experimental HERFD-XAS spectrum of the $Sc_xGa_{1-x}N$ sample ($x = 0.059$) with the simulated XANES spectrum of the $Sc_1Ga_{15}N_{16}$ supercell, corresponding to $x = 0.0625$. The figure shows spectra obtained with (a) $E \perp c$ and (b) with $E \parallel c$.

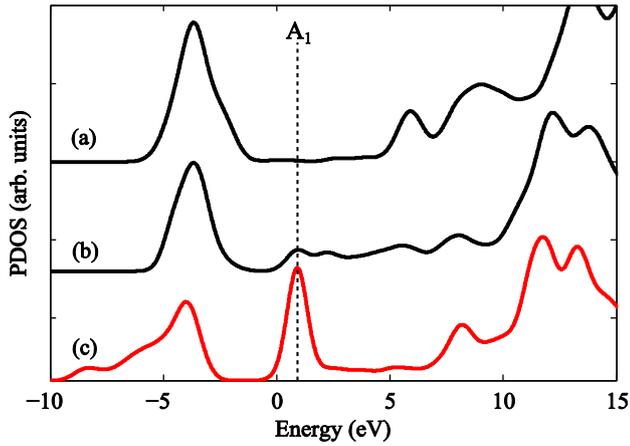

Fig. 7 Broadened DFT-calculated unoccupied $p$ PDOS for Sc bonded to N in different local coordination environments. (a) octahedral ($O_h$), (b) trigonal-bipyramidal ($B_k$), (c) tetrahedral ($T_d$), as in $Sc_1Ga_{15}N_{16}$. The $A_1$ pre-edge peak position is indicated by the dashed line.

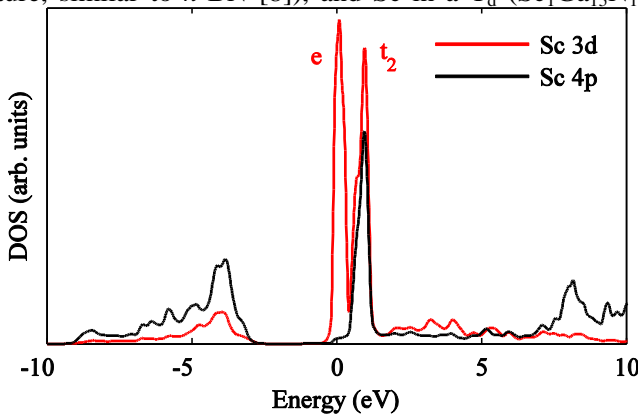

Fig. 8 Projected unoccupied density of states for Sc $3d$ and $4p$ states in distorted wurtzite $Sc_1Ga_{15}N_{16}$.

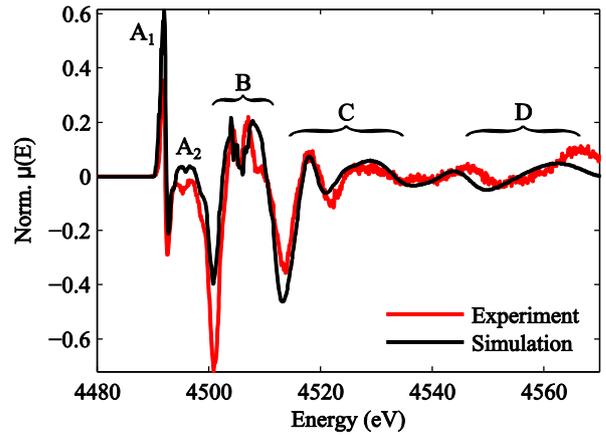

Fig. 9 Comparison of experimental X-ray linear dichroism (XLD) for $Sc_xGa_{1-x}N$ with $x = 0.059$ and theoretical XLD spectra for $Sc_1Ga_{15}N_{16}$ supercells calculated by FDMNES.

As expected from symmetry, the PDOS calculations of the $O_h$ bonding environment clearly show the absence of any pre-edge peak features, and only very weak features are visible in the $B_k$ five-fold coordinated structure.

The $A_1$ peak can be interpreted as arising from transitions to the Sc mixed $3d$–$4p$ valence states, as can be seen in Fig. 8. In particular, the PDOS calculations indicate that the Sc $4p$ states overlap only with the $t_2$ $3d$ states and not the $e$ $3d$ states, as expected from symmetry considerations, which explains the presence of only a single $A_1$ peak. Accurately reproducing the intensity of pre-edge peaks theoretically poses a significant challenge, due to the large number of parameters that affect it: the local bonding symmetry (averaged over all atoms), the level of theory employed (pre-edge features may present multiplet effects not taken into account in the current simulations) and the final convolution. Thus, we cannot use the observed reduction in the $A_1$ peak intensity quantify the fraction of Sc going out of perfect local $T_d$ symmetry.

The X-ray natural linear dichroism, XLD, in XAS [31] is obtained by taking the difference between the XAS spectra with the electric field vector perpendicular (Fig. 6a) and parallel (Fig. 6b) to the crystallographic $c$ axis. XLD is only dependent on the crystal symmetry and its amplitude, compared to the calculated one, can be used to extract a quantitative parameter of the level of substitutional inclusion [32]. In Fig. 9 we report a comparison of the experimental and theoretical XLD spectra. A good agreement in both shape and intensity of the spectral features is visible. Considering the strongest dichroism of the main absorption edge (B) for all three compositions, it yields a quality factor above 95% which confirms the previous EXAFS results.

## 4. Conclusions

In this paper, we have analysed the local bonding and electronic structure of Sc in dilute $Sc_xGa_{1-x}N$ films grown epitaxially on GaN. We find that Sc substitutes for Ga in a slightly distorted wurtzite structure for $x \leq 0.059$, with a local increase in the $u$ parameter around the Sc atoms, as determined from EXAFS and HERFD-XAS measurements. An intense pre-edge peak was observed in the Sc $K_\alpha$ HERFD-XAS spectra, which is typical for $d^0$ transition metal atoms. This pre-edge peak is reproduced best by DFT simulations based on relaxed $Sc_xGa_{1-x}N$ supercells in which a local deviation away from perfect $T_d$ bonding symmetry arises around Sc atoms, associated with an increased local $u$ value at the Sc atoms.

We conclude from both EXAFS and HERFD-XAS analysis that, for $x \leq 0.059$, at least 95 ± 5% of Sc atoms incorporate substitutionally for Ga atoms in a nearly-perfect $T_d$ coordination environment. The remaining fraction of Sc is likely to be located at planar defects such as inclusions of the cubic phase, or stacking mismatch boundaries in the films, where the local bonding environment is different. In this compositional range, epitaxial $Sc_xGa_{1-x}N$ films retain the wurtzite structure of the GaN substrate.

The close correspondence between the XANES spectra simulated by FDMNES and the features in the experimental HERFD-XAS data indicate that our predictions regarding the crystal and electronic structure of $Sc_xGa_{1-x}N$ are correct in the dilute regime up to compositions of at least $x = 0.059$ in which the wurtzite structure is retained [6]. These results give us the confidence that related predictions, such as the retention of wide, direct band gaps for phase-stable wurtzite $Sc_xGa_{1-x}N$ up to at least $x = 0.27$, are also correct. However, further experimental studies of higher Sc content films is required before integration into III-nitride devices will be possible.

## Acknowledgements

MAM acknowledges support from the Royal Society through a University Research Fellowship and through the ERC Starting Grant 'SCOPE'. The XAS experiments were performed at the European Synchrotron Radiation Facility via proposal HE-3947.